\documentclass[a4paper,10pt]{article}
\usepackage{graphics}

\usepackage{url}
\usepackage{graphicx}
\usepackage{color}
\usepackage{tabularx}
\begin{document}
\title{Atlas-Based Prostate Segmentation Using an Hybrid Registration}
\title{Atlas-Based Prostate Segmentation Using an Hybrid Registration}

\author{
\parbox{0.8\textwidth}{\centering
S\'ebastien Martin (1) Vincent Daanen (2) Jocelyne Troccaz (1)\\
\small
(1) Laboratoire TIMC-IMAG, \'equipe GMCAO, CNRS-UJF, Institut d'Ing\'enierie de l'Information de Sant\'e, 38706 La Tronche cedex, France \\
(2) KOELIS, 5 avenue du Grand Sablon, 38700 La Tronche, FRANCE 
}
}

\maketitle
\begin{abstract} 
\textit{Purpose:} This paper presents the preliminary results of a semi-automatic method for prostate segmentation of Magnetic Resonance Images (MRI) which aims to be incorporated in a navigation system for prostate brachytherapy. 

\textit{Methods:} The method is based on the registration of an anatomical atlas computed from a population of 18 MRI exams onto a patient image. An hybrid registration framework which couples an intensity-based registration with a robust point-matching algorithm is used for both atlas building and atlas registration.  

\textit{Results:} The method has been validated on the same dataset that the one used to construct the atlas using the "leave-one-out method". Results gives a mean error of 3.39 mm and a standard deviation of 1.95 mm with respect to expert segmentations.

\textit{Conclusions:} We think that this segmentation tool may be a very valuable help to the clinician for routine quantitative image exploitation. 


\end{abstract}

\section{Introduction}
\label{intro}
Adenocarcinoma of the prostate is the most common cancer expected to occur in men in 2008 \footnote{Excluding basal and squamous cell skin cancers and in situ carcinoma except urinary bladder}. A total of 186,320 new cancer cases and 28,660 deaths from prostate cancer are projected to occur in the United States in 2008 \cite{Jemal2008}.

Magnetic Resonance Imaging (MRI) allows the detection of the prostate and structures therein with higher accuracy than ultrasound images (US). Automatic segmentations of the prostate in MRI have shown to be better correlated to manual expert segmentations than automatic segmentations in US. In order to make dose planning easier in prostate brachytherapy, we introduced a MRI/US registration method based on contours (\cite{Reynier_2004} \cite{Daneen2006}); in the current version, segmentations are produced manually which is a practical limitation. The objective is to automate this segmentation phase; in a first stage we focused on MRI data. For practical use in a clinical situation, the algorithm must be accurate, relatively fast, and must be limited to as simple interaction with the expert as possible.

Works on automatic or semi-automatic techniques for segmentation of the prostate in MRI are limited. Zwiggelaar and al. \cite{Zwi2003} have developed a technique based on polar-transform. Zhu and al. used an Active Shape Model to achieve a semi-automatic segmentation. Zhu and al. improve their previous work by using a combined 2D/3D Active Shape Model \cite{Zhu2004} \cite{Zhu2005}. \cite{Klein2008} achieve an automatic segmentation using an atlas-based method. Other works deal with CT or US modalities~: \cite{costa2007} use a coupled bladder and prostate segmentation in CT images to achieve an automatic prostate segmentation. \cite{Hodge2006} use an 2D active shape model to segment prostate in Ultrasound images. Our proposed segmentation method is based on the registration of an anatomical atlas to an individual image. The original aspect of this work consists to use a hybrid registration algorithm for atlas building and atlas matching.

\section{Material}
\label{material}
MRI acquisitions have been realized with a trans-rectal probe. Volumes have been acquired with a voxel size of $0.53 \times 0.53 \times 3.12$ $\textrm{mm}^3$ and have been resampled to a voxel size of $1.0 \times 1.0 \times 1.0$ $\textrm{mm}^3$. The set of data used for model building is presently composed of 18 MRI exams. All acquisitions have been realized on patients where a prostate cancer was diagnosed, before a brachytherapy treatment.
One expert has segmented all the MRI exams. These segmentations are used for model building and evaluation.
\section{Method Overview}
\begin{figure}
\caption{Overview of the method}
\begin{center}
\includegraphics[width=10cm]{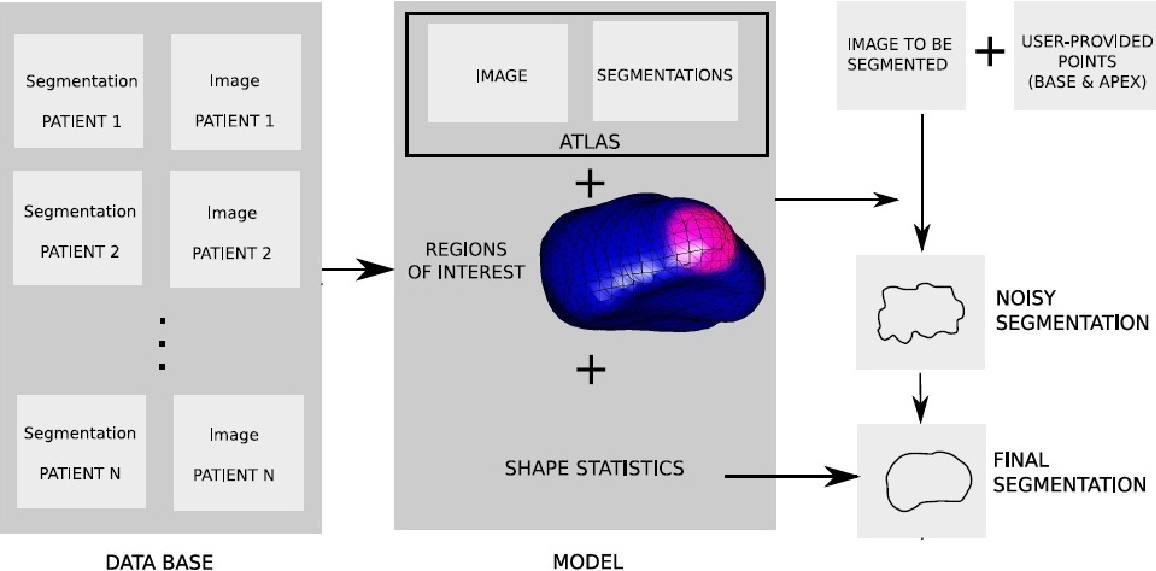}
\end{center}
\label{method_overview}
\end{figure}
\label{overview}
The proposed segmentation method is based on the registration of an anatomical atlas to an individual image. We both use an hybrid registration method for atlas matching and atlas construction. A shape model of the prostate constructed from the population used to build the atlas is used to regularize segmentations obtained after the atlas matching. Figure \ref{method_overview} illustrates the principle of the method.

The hybrid registration which exploits image and geometric information is based on two alternated registrations. The framework registers a template image $T(x)$ onto a reference image $R(x)$ and a set of points $M = \left\lbrace m_0, ..., m_N\right\rbrace $ (called "model points" and belonging to the template image) to a set of points $S=\left\lbrace s_0, ...,s_M\right\rbrace $ (called "scene points" and belonging to the reference image). The description of the general framework is done in section \ref{hybridRegistration}. Figure \ref{fig::scheme} illustrates the principle of the registration.

During the anatomical atlas construction described in section \ref{atlas_construction} we use the hybrid registration framework to map all individuals of the database onto a common reference by exploiting an image similarity measure and a distance between expert segmentations.

The same framework is also used during the atlas-based segmentation to incorporate a user interaction through a robust point to surface matching (see section \ref{atlas_to_subject}). The interaction is realized by asking the user to select one point on two specified regions in the image to be segmented. These regions are the base and the apex of the prostate and are defined by assigning a prior probability of match to each model point embedded in the atlas. This is described in section \ref{interest_points}.

After the atlas-based segmentation a projection of the resulting segmentation on the shape space is done to regularize results (section \ref{shape_model}). A brief description of shape model construction is given in section \ref{shape_model_projection}.

\section {An Hybrid Registration method}
\subsection {A dual energy minimization}
\label{hybridRegistration}
\begin{figure}[h]
\begin{center}
\includegraphics[width=10cm]{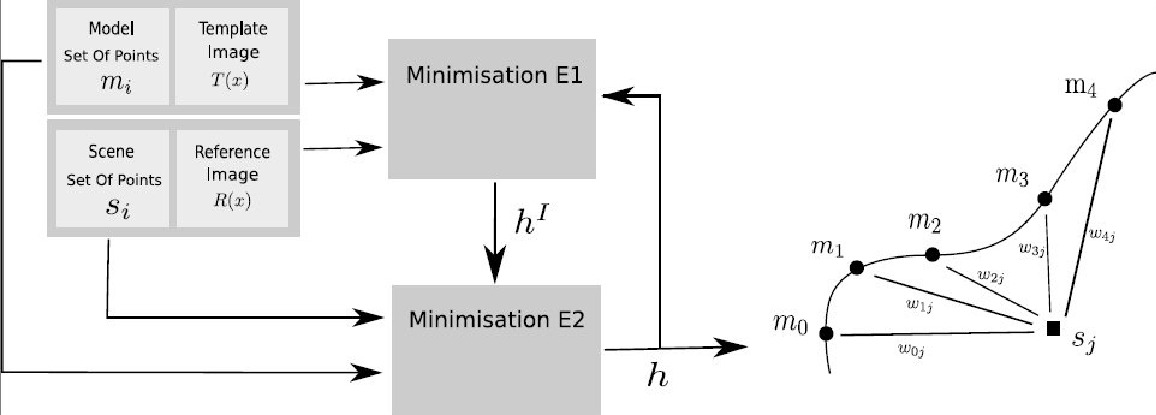}
\end{center}
\caption{\textbf{Hybrid registration}; left: principle of hybrid registration; right: correspondences between model points (circles) and a scene point (square) are modelled by assigning weights $w_{ij}$ to each couple ($m_i$, $s_j$). }
\label{fig::scheme}
\end{figure}
Coupling intensity and geometric information can be very interesting to improve the quality of registration algorithms. Area of applications are for example the incorporation of segmentations extracted by some image processing techniques into an intensity-based registration (\cite{hellier2003}), or the introduction of expert constraints to an intensity-based registration (\cite{azar2006}). 

Here we propose a method that alternates the minimization of two criteria to both take into account an intensity-based distance and a geometric distance. The minimization of the intensity-based distance aims at matching the template image $T(h(x))$ on the reference image $R(x)$ while the minimization of the geometric distance aims at matching model points $m_i$ belonging to the template image to scene points $s_j$ belonging to the reference images. The two criteria are incorporated into two dual energies functions which are minimized alternatively through two registrations algorithms:
\begin{eqnarray}
\label{eq::E_im} 
E_1 &=& E_{sim}(T, R, h^I) \\[8pt]
\label{eq::E_fus}
E_2 &=& \sum_j \sum_i w_{ij} \Vert h^{-1}(m_i)-s_j \Vert^2  + \beta \Vert h^I - h\Vert^2
\end{eqnarray}
Where $h^I$ and $h$  are two vector fields, $E_{sim}$ is a similarity measure between two images and $\Vert . \Vert$ is a distance between two vector fields (L2 distance). $h^I$ estimates intensity matches between $T(x)$ and $R(x)$. $h$ is a trade-off between geometric constraints and intensity matches. (See figure \ref{fig::scheme}). 

The hybrid registration operates as follows~:
\begin{enumerate}
\item initialize $h(x)$ and $h^I(x)$ to identity.
\item minimize $E_1$ w.r.t $h^I$ using the registration algorithm of \ref{regE1}
\item minimize $E_2$ w.r.t $h$ using the registration algorithm of \ref{regE2}
\item set $h_I = h$ 
\item \textsc{if convergence reached stop else goto 2}
\end{enumerate}
The intensity information is introduced in the first term of eq. (\ref{eq::E_im}) by the similarity energy $E_{sim}$. 

The geometric constraint is introduced by the first term of eq. (\ref{eq::E_fus}) which is a geometric distance between model points $m_i^t = h_t^{-1}(m_i)$ (at iteration $t$) and scene points $s_j$. The weight $w_{ij}$ determines the correspondence between the two point sets and can be estimated by several criteria such as ICP or softAssign \cite{Rangarajan}. (see figure \ref{fig::scheme}) 
\subsection {Registration based on the minimization of $E_1$}
\label{regE1}
Physically-based models are often used to ensure the coherence of the deformation field. In an elastic model, elastic forces are proportional to the displacement field while in a fluid model, fluid forces are proportional to the rate of change of the deformation field. The fluid regularization allows the estimation of large deformations while maintaining the topology of the deformation field \cite{thirion98}. An elastic model does not guarantee to keep the coherence of the deformation field. However fluid models are often under-constrained and the resulting deformation field can be non-satisfactory. In our implementation we both use an elastic model and a fluid regularizer (\cite{stefanescu2003}). 

Our method implements two levels of regularization and the registration is done as follows~:\\
\textsc{Do Until Convergence}
\begin{enumerate}
\item Compute $\nabla_{p(x)} E_1[h^I(x) \circ (x+p(x)) ] = \nabla_{c} E_1$.
\item Compute a correction~: $c(x) = k \times \nabla_{c} E_1 \star G_{\sigma}(x)$. Where $k$ is a scalar controlling the magnitude of the vector field $c(x)$ and $G_{\sigma}(x)$ is a Gaussian kernel of standard deviation $\sigma$.
\item Compose the correction field to the current deformation.\\ 
	$h_n^I(x) = h_{n-1}^I(x) \circ (x+c(x))$ 
\item Regularize the deformation field $h^I(x)$ by decreasing $E_{elas}$ (elastic regularization)
\end{enumerate}
where $E_{elas}$ is the linear elasticity potential~:
\begin{eqnarray}
E_{elas} = \int \frac{\lambda+\mu}{2} \Vert \nabla.u^I \Vert^2 + \frac{\mu}{2} \sum_i^3 \Vert \nabla u^I_i \Vert^2
\end{eqnarray}
where $h^I(x) = x + u^I(x)$ and $u^I_i$ denotes the $i^{th}$ component of $u^I(x)$. 
% 1ere étape
During the first step, following (\cite{trouve1998}, \cite{stefanescu2003}), we look for a small perturbation $p(x)$ that minimizes $E[h(x) \circ (x+p(x)) ]$ and call it $\nabla_{c} E_1$.
% 2eme étape
In a second step, we compute a correction field by taking a fraction of the smoothed gradient $\nabla_{c} E_1$. The smoothing allows the filtering of noisy values of the gradient and allows to ensure the invertibility of the deformation field. This smoothing can be related to a fluid regularization controled by the standard deviation of the Gaussian kernel ($\sigma$). The scalar $k$ is tuned in order that the maximum magnitude of $c(x)$ is lower than 0.5 voxel.
% 3eme étape
During the step 3 the correction field $c(x)$ is composed to the current deformation field. 
% 3eme étape
During the last step a regularization of the deformation field is achieved by the minimization of $E_{elas}$. This minimization results to a classical diffusion equation which is solved using a Gauss Seidel algorithm and a semi implicite scheme \cite{Press96}. The diffusion time tunes the elastic registration.
\subsection {Registration based on the minimization of $E_2$}
\label{regE2}
For the minimization of energy $E_2$, the algorithm uses a family of tensor product splines to model the deformation field.
\begin{eqnarray}
h(x,p) = x + \sum_{k,l,m} u_{kml} B_k(x_0) B_l(x_1) B_m(x_2)
\end{eqnarray}
$u_{klm}$ are the spline deformation coefficients which comprise a parameter vector $p$. $B_k$, $B_l$, $B_m$ are the spline basis functions. A regularization is used to put prior constraints on spline parameters. The cost function to be minimized becomes:
\begin{eqnarray}
\label{eq::spline_cost} 
C(p) = E_2( h(x,p) ) + \int \sum_i^3 \Vert \nabla u_i \Vert^2
\end{eqnarray}
where $h(x) = x + u(x)$ and $u_i(x)$ denotes the $i^{th}$ component of $u(x)$. The approximation of the integral is based on finite differences and the minimization of $C(p)$ (\ref{eq::spline_cost}) is done by gradient descent. 
%The derivative of the first term of eq \ref{eq::E_fus} w.r.t $p_i$ ($i^{th}$ component of $p$) is~:
%\begin{eqnarray}
%\partial_{p_i} &&\sum_j \sum_i w_{ij} \Vert h^{-1}(m_i)-s_j \Vert^2 \\&=& \sum_j \sum_i 2 w_{ij} (\partial_{p_i} h)^T \times [ \nabla h^{-1}(h(m_i^t)) ] \times (m_i^t-s_j) \\
%&=& \sum_j \sum_i 2 w_{ij} (\partial_{p_i} h)^T \times [\nabla h(m_i^t)]^{-1} \times (m_i^t-s_j) 
%\end{eqnarray}
%Other derivatives are straightforward.
\subsection{Inversion method for computing $m_i^t$}
\label{inversion_method}
At each iteration, model points evolves according to the backward transformation $h^{-1}(x)$. We denote by $m_i=m_i^0$ the initial position of surface points (at iteration 0). $h^{-1}(m_i) = m_i^t$ is the transformed point at iteration $t$. 
The inversion is realized for each point $m_i$, by minimizing the criterion $f(p) = \Vert h(p)-m_i \Vert^2$ with respect to $p$. The minima of $f(p)$ is reached for $p=h^{-1}(m_i)=m_i^t$. We start the minimization by setting $p=m_i^{t-1}$.
\section {Model Building}
\label{model_building}
As introduced previously, the model includes~:
\begin{itemize}
\item the atlas (the mean image and associated contours)
\item a shape model of the prostate embedded in the atlas
\item prior probabilities of match to user provided points (base and apex)
\end{itemize}
\subsection{Atlas construction}
\label{atlas_construction}
\begin{figure}
\caption{Example of atlas generated using only intensity information (left) and using intensity and geometric information (right)}
\begin{center}
\includegraphics[width=9cm]{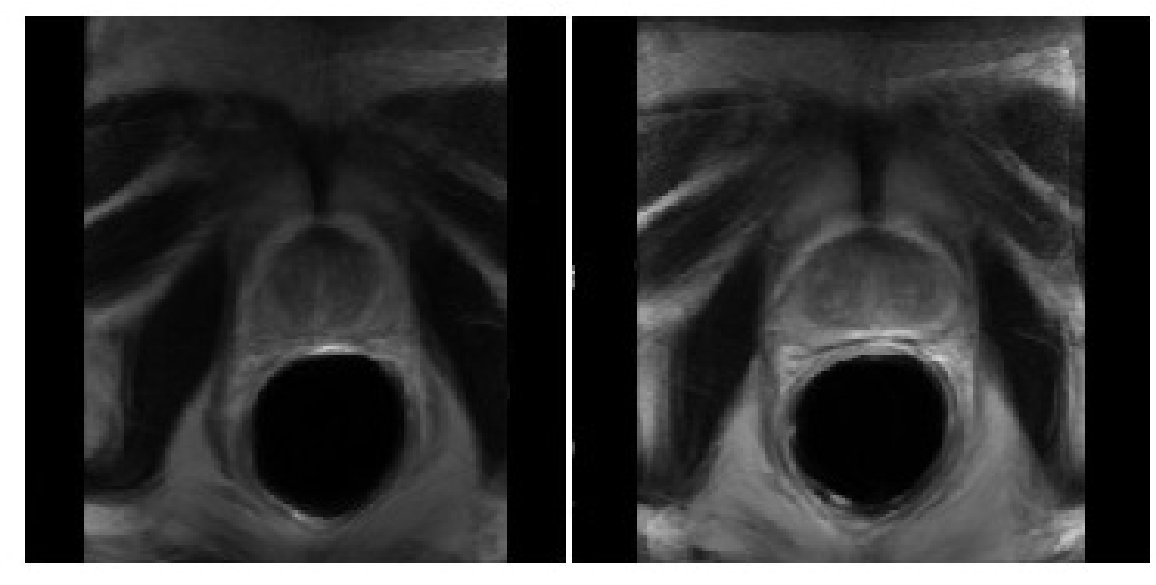}
\end{center}
\end{figure} 
Atlas-based segmentation has become a standard approach to organ delineation for many fields of medical image analysis particularly for the study of the brain (\cite{Klein2008}). In these methods a template (atlas) is registered to a new patient image. The spatial transformation obtained from the registration process is used to map anatomical segmentations of the atlas onto the new subject. The main issue of these methods is the template selection. A common approach to construct an atlas is to label a particular image (Talairach atlas in neurosurgery for instance). Another widely used method is to construct an anatomical atlas from a population by averaging multiple co-registered images. Our method constructs an anatomical atlas by registering a population to a reference $R$ manually chosen among the population. The reference was selected to be close to the mean prostate size. First, for each image $I_i$ of the population an inhomogeneity correction is performed. Then, all individuals are elastically co-registered to the reference $R$. 

After registrations, we have a set of transformations $T_i$ that map the reference onto each individual $I_i$. The atlas image is obtained by averaging the registered images $I_i \circ T_i$ and the atlas contour (prostate) is obtained by taking the segmentation of the reference. Many works use an averaging of the registered segmentations $S_i \circ T_i$ to obtain the mean segmentation (where $S_i$ denotes the segmentation of the image $i$) or a probability map of the atlas structures \cite{M2004} \cite{H2005}. We think that these methods are more interesting when the registration of the atlas is used as a prior information for another segmentation method. This will not improve the atlas contour in our case. The atlas construction method described is then iterated by taking the current atlas image as a new reference; we restart all registrations and the atlas construction until the mean image does not change significantly.

Very often, anatomical atlas construction only makes use of intensity information to estimate the transformation to the reference. However, such methods are not usable in our case because of too important mismatches which may appear between reference segmentation and deformed template segmentation. The construction of an atlas using this method gives poor results. To address this problem we use the hybrid method introduced in section \ref{hybridRegistration}. The  geometric criterion (ICP) works on the contours given by the expert segmentations while the intensity based criterion (SSD) works on the images using a SSD as a similarity measure.  The matching problem is then solved using an ICP criterion (``hard assignment''): for each scene points $s_j$, $w_{ij}=1$ if $m_i$ is the closest model point of $s_j$ else $w_{ij} =0$.
\subsection{Shape Model}
\label{shape_model}
Our method incorporates statistical shape information in order to improve the quality of the resulting segmentations \cite{cootes1994}. When computing a statistical shape model, a fundamental problem is the determination of a set of corresponding points between instances (cloud of points) of the population. In the previous section, we determined a set of correspondences (transformations) which maps the reference onto each individual, based on a framework which couples geometric and intensity based registration. These dense transformations are now used to generate a set of corresponding points by deforming the cloud of points of the reference onto each individual. The resulting sets of corresponding points are statistically analyzed by principal component analysis to build a shape model.
\subsection {Region of interest (Base and Apex)}
\label{interest_points}
To guide a registration algorithm, one can couple landmark and intensity-based registration (\cite{azar2006}). This is very useful in presence of large deformations to forbid the algorithm to fall into local minima or to incorporate a strong knowledge (expert) when image information is poor. However the determination of landmarks is often very subjective. In our method instead of using landmarks we prefer to model the correspondence between a user-provided point and a point which belongs to the surface in a probabilistic framework. We defined two interest areas on the prostate (base, apex) by assigning a prior probability (of correspondence with the user provided seed) to each atlas model point. The expert user is then asked to determine one point on each region. The correspondence between the user provided point in the study image and the prostate surface in the atlas volume is then carried out iteratively in a probabilistic way as described in section (\ref{atlas_to_subject}).

A simple method to assign prior probabilities of matches ($\pi_{ij}$) to each couple $(m_i, s_j)$ =(model point, user-provided point) is to manually define a point $d_j$ in a center of the region $j$ and to assign probabilities according to the distance to this point using a shape function $G$ (for example a Gaussian kernel). The denominator of eq (\ref{eq::assign_prior}) is used for normalization~:
\begin{eqnarray}
\label{eq::assign_prior}
\pi_{ij} = \frac{G(m_i-d_j)}{\sum_i G(m_i-d_j)} 
\end{eqnarray}
\section{Prostate segmentation}
\subsection{Anatomical Atlas To Subject Registration}
\label{atlas_to_subject}
\begin{figure}
\label{fig::results_seg}
\begin{center}
\includegraphics[width=10cm]{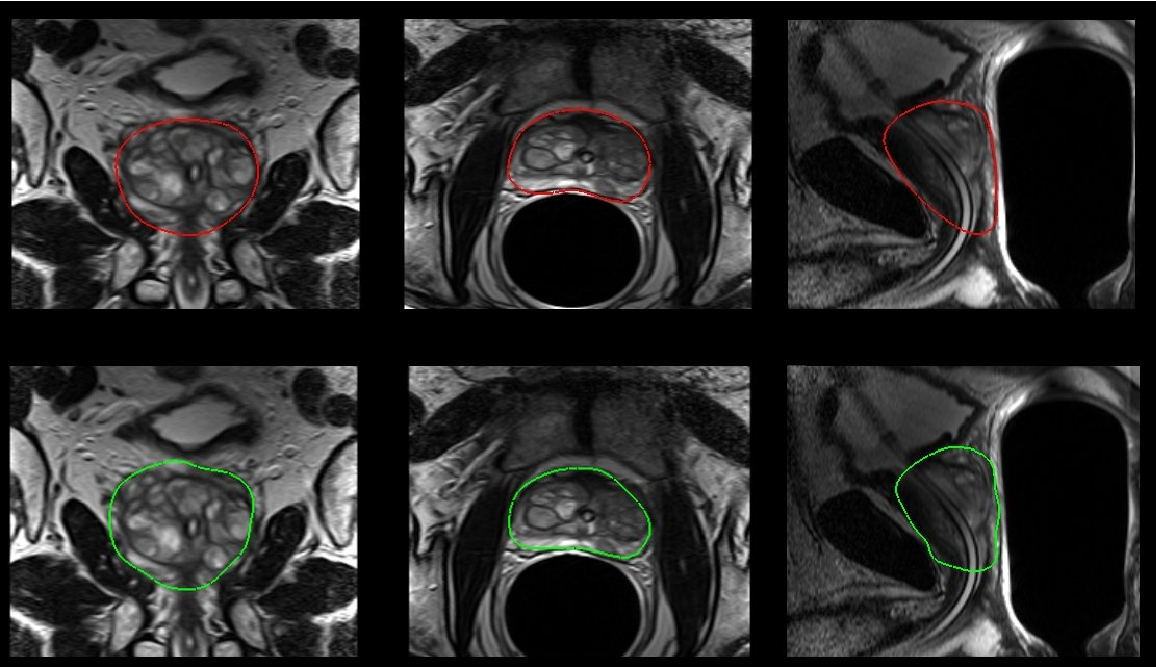}
\end{center}
\caption{Examples of expert and automatic segmentations : top row expert segmentations, bottom row automatic segmentations; from the left to the right coronal slice, transverse slice, sagittal slice}
\end{figure} 
The algorithm starts with two interest points (user-provided points) and corresponding prior probability of match defined for each atlas model points. The first step consists in rigidly registering the atlas to the patient image using an intensity-based registration method using the SSD as similarity measure. In a second step we use the hybrid registration algorithm for the non rigid registration of the atlas onto the study image. The similarity measure used in the hybrid registration is also a SSD. The model points $M^t = \left\lbrace m_0,...,m_N \right\rbrace $ represent the prostate surface embedded in the atlas. The scene points $S=\left\lbrace s_0,...,s_M\right\rbrace $  are user provided points. We recall that $m_i^t = h_t^{-1}(m_i)$ is the moving model point at iteration $t$ and $\pi_{ij}$ be the prior probability of match of the scene point $s_j$ to the model point $m_i^t$.
\begin{figure}
\label{fig::results_seg2}
\begin{center}
\includegraphics[width=10cm]{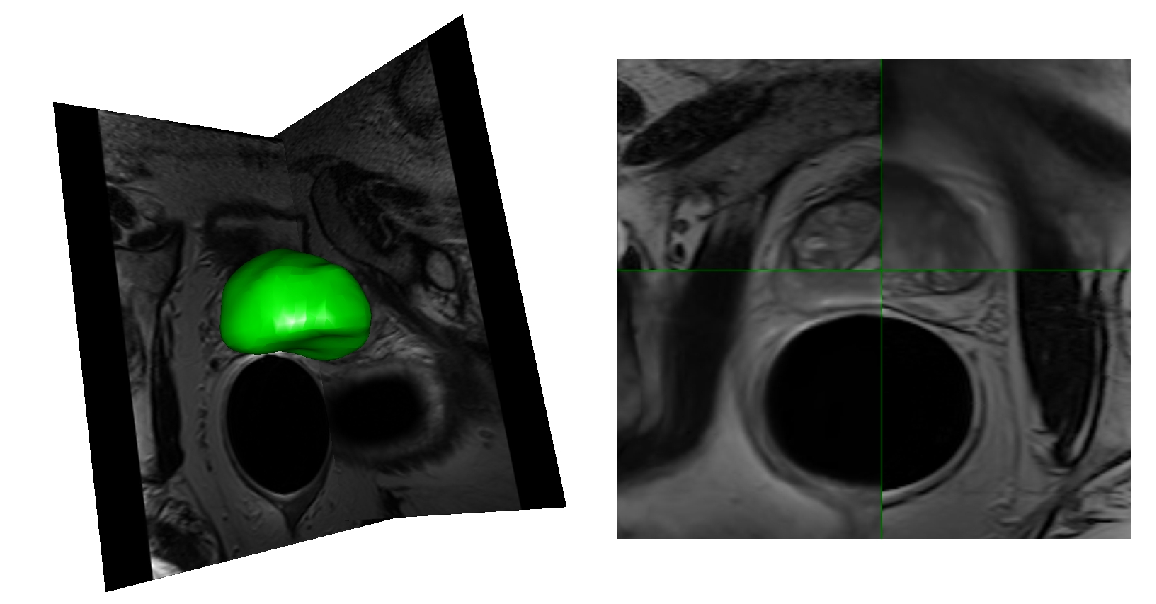}
\end{center}
\caption{Automatic segmentation -- deformated atlas ; from the left to the right~: 3D view of automatic segmentations, fusion of deformed atlas (top right / bottom left) and study image (top left / bottom right)}
\end{figure} 
\subsubsection {Geometric Criterion}
Here we present the method used to determine correspondence between user-provided ($s_j$) points and model points ($m_i$ ; prostate) using prior probabilities of match (regions of interest) defined in section (\ref{interest_points}). The framework is based on the soft assign method \cite{Rangarajan}. We now consider a ``binary random match matrix'' $W_{ij}$ which randomly associates the scene point $s_j$ to model point $m_i^t$ (if $W_{ij}=1$ and other elements of the row are set to zero, this means that the point $s_j$ corresponds to the point $m_i^t$). $\pi_{ij}$ defined in section \ref{interest_points}) represents a prior probability of match $P(W_{ij}=1)$ associated to each couple of points $(m_i^t,s_j)$. 
We now model the observation law (likelihood) as a Gaussian process. if a model point $m_i^t$ corresponds exactly to a scene point $s_j$, the likelihood of $s_j$ is:
\begin{eqnarray}
P(s_j|W_{ij}=1) \propto \exp \left\lbrace -\frac{\Vert s_j - m_i^t \Vert^2}{\sigma_m} \right\rbrace 
\end{eqnarray}
Weights $w_{ij}$ introduced in eq. (\ref{eq::E_fus}) are determined by computing $P(W_{ij}=1| s_j)$~:
\begin{eqnarray}
P(W_{ij}=1|s_j) &=& \frac{\pi_{ij} \exp \left\lbrace - \frac{\Vert s_j - m_i^t \Vert^2}{\sigma_m} \right\rbrace  }
	{\sum_i^N \pi_{ij} \exp \left\lbrace  - \frac{\Vert s_j-m_i^t \Vert^2}{\sigma_m} \right\rbrace  } \\
	&=& w_{ij}
\end{eqnarray}
These weights are recomputed at each iteration of the minimisation of $E_2$ by the algorithm presented in section \ref{regE2}. 
\subsection {Shape Model Projection}
\label{shape_model_projection}
A shape model of the prostate has been built from a set of corresponding points obtained by deforming the mesh of the reference onto each individual (using transformations obtained during the atlas construction). At the end of the atlas construction the reference is the atlas. It results that we have a shape model of the prostate surface which is embedded in the atlas. During the atlas to study matching the atlas mesh is deformed to segment the prostate on the study image. To regularize this segmentation which can be noisy, we can now compute the most probable shape according to the shape statistics. To this end, we estimate modes of variation and then we apply constraints to modes, to ensure plausible shapes. This is done by limiting their absolute value to be less than $3 \sqrt \lambda_i$. ($\lambda_i$ is the eigen-value of mode $i$)
\section {Results}
\label{results}

\begin{table}
\caption{\sc{Volume Based Metrics; Sensitivity (SENS) and Positive Predictive Value (PPV) with respect to one expert segmentation} }
\begin{center}
% use packages: array

\begin{tabular}{ccc}
\begin{tabular}{c p{1.8cm} p{1.8cm}}
\hline
\sc{Image} & \sc{ Sens } & \sc{ PPV } \\ 
\hline \hline
1 & 0.91 & 0.75  \\ \hline 
2 & 0.85 & 0.96	 \\ \hline
3 & 0.95 & 0.6  \\ \hline
4 & 0.92 & 0.86  	\\ \hline
5 & 0.98 & 0.73  	\\ \hline
6 & 0.94 & 0.76  	\\ \hline
7 & 0.87 & 0.47 	\\ \hline
8 & 0.84 & 0.94   	\\ \hline
\\ 
\end{tabular}
&  &
\begin{tabular}{c p{1.8cm} p{1.8cm}}
\hline
\sc{Image} & \sc{ Sens } & \sc{ PPV } \\ 
\hline \hline
9 & 0.91 & 0.79  	\\ \hline 
10 & 0.81 & 0.77  	\\ \hline
11 & 0.94 & 0.75  	\\ \hline
12 & 0.87 & 0.78  	\\ \hline
13 & 0.89 & 0.79  	\\ \hline
14 & 0.85 & 0.84  	\\ \hline
15 & 0.87 & 0.87  	\\ \hline
16 & 0.77 & 0.83  	\\ \hline
17 & 0.94 & 0.72  	\\ \hline
\end{tabular}
\end{tabular}
\end{center}
\label{tb::area_metric}
\end{table}

\begin{table}
\caption{ \sc{Distance Based Metrics (the mean size of the prostate in the axis base/apex is about 48} \textit{mm})}
\begin{center}
% use packages: array
\begin{tabular}{l l|l l|l l | l l}
\hline
Apex &  & Central Zone & × & Base & × & All & \\ \hline  
Mean (mm) & STD   & Mean (mm) & STD    
	& Mean (mm) & STD & Mean (mm) & STD \\  \hline \hline
2.91 & 1.34 & 2.4 & 1.29 & 4.30 & 2.00 & 3.39 & 1.95\\ \hline
\end{tabular}
\end{center}
\label{tb::distance_metric}
\end{table}
The method has been tested with the same dataset than the one used for atlas and shape model construction. The ``leave-one-out'' method has been used for this purpose (i.e.: the evaluated data -- the patient to be automatically segmented -- is removed from the set of data used for model construction). The reference image used for atlas building is not used for validation to avoid additionnal bias when choosing a new reference; it results that we have 17 patients for tests.

Results obtained using a volume-based metrics are presented in Table \ref{tb::area_metric}. The two distances used are the ``sensitivity'' and the ``positive predictive'' value defined as follows~:
\begin{center}
% use packages: array
\begin{tabular}{lll}
$ PPV = \frac{TP}{TP+FP} $ & $ SENS = \frac{TP}{TP+FN} $ \\ 
\end{tabular}
\end{center}
Where $TP$ denotes the number of true positives, $FP$ denotes the number of false negatives and $FP$ denotes the number of false positives. Results obtained using a distance based metrics are presented in Table \ref{tb::distance_metric}. We have divided the prostate in 3 zones to show how the algorithm performs spatially. Results show that the base of the prostate is more difficult to segment. The accuracy in the apex zone and in the central zone is good. The convergence of the registration algorithm in the base of the prostate is difficult due to the large interindividual variability observed. For small prostates (volume less than 25 cc for a mean volume of 41 cc) the algorithm behaves less accurately (mean error of about 5.5 mm in 3 cases). We are very confident in our ability to solve this problem using several atlases depending on the size of the prostate in the study image. 
\section {Conclusion}
\label{conclusion}
% sum up (atlas-based segmentation using an hybrid framework)
% 1) hybrid registration = construction + atlas based seg
This paper has presented an atlas-based semi-automatic prostate segmentation method for MRI. The method makes use of an hybrid registration framework which can deal with geometric and image-based information for both atlas construction and atlas-based segmentation.
% result
Results are very promising expect for small prostates (volume less than 25 cc for a mean volume of 41 cc) where the convergence is very difficult (mean error of about 5.5 mm in 3 cases). The use of an additional atlas built from a population of small prostates will certainly help to solve the problem.
% 2) population based shape model construct is used to regularized estimated seg
The hybrid registration allows for an efficient estimation of correspondences across individuals of the database. These correspondences are used for atlas construction and for shape model building.
% 3) beaucoup de works sur fusion iconique landmarks (probleme de hard correspondance)
Many works use landmarks to introduce geometric constraints in  intensity-based registration methods \cite{hellier2003} \cite{azar2006}. However the frequent limitation of these approaches is the determination of hard correspondences between reference and template image which is often very subjective. 
% 4) ce papier s'affrenchie de ces limitation en fusionant une approche iconique et geometric avec estimation des correspondance
This work overcomes these limitations by iteratively estimating correspondences between the surface embedded in the atlas, which deforms during the registration and few user-provided points.
% 5) le probleme des correspondance est resolu de facon elégante grâce à la définition de region d'interet l'interaction utilisateur est simple et permet de contraindre le recalage dans des regions critiques (base appex) ou l'information image est pauvre.
Correspondences are estimated in an elegant way using a probabilistic framework and prior information on expected correspondences between user-provided points and prostate surface.
% 6) Bien que le critere gemetrique est limité à un matching point utilisateur surface il serait tout a fait possile d'utiliser des points extrait grace à des algorithmes de traitement d'images et de les introduire dans le framework.
An extension of this work could be to use additional geometric features automatically extracted from the image by some image processing techniques directly in the hybrid registration framework. 
Current work deals with the addition of other expert segmentations into the database and with qualitative evaluation of the method from the clinical side.
% 8) d'autre organes peuvent être segmenter en utilisant le meme principe et peuvent servir de primitives gemotrique à un recalage surface-echo direct ou basé sur des contours ce qui peut être une aide pour la segmentation en echo (dans le contexte de curiethérapie)
%Other organs in the neighborhood of the prostate can be delineated using the same method (muscle, rectum, pubic bone) and could serve in a MRI/US registration which could be based on a surface to image registration \cite{wu2003} or on a surface to geometric feature registration \cite{Daneen2006}. 
%
% BibTeX users please use

\bibliographystyle{splncs}
\bibliography{IJCARS_smartin}

\end{document}